\documentclass[a4paper,fleqn]{cas-dc}

\usepackage[authoryear,longnamesfirst]{natbib}

\def\tsc#1{\csdef{#1}{\textsc{\lowercase{#1}}\xspace}}
\tsc{WGM}
\tsc{QE}
\newtheorem{theorem}{Theorem}
\usepackage{algorithmic}
\usepackage{algorithm}

\begin{document}
\let\WriteBookmarks\relax
\def\floatpagepagefraction{1}
\def\textpagefraction{.001}

\shorttitle{DDC of TSR deployment with Learned Dynamics}    

\shortauthors{A.Jin,F.Zhang,P.Huang}  

\title [mode = title]{Data-Driven Optimal Control of Tethered Space Robot Deployment with Learning Based Koopman Operator}  


%\tnotetext[1]{Title footnote text} 

\author[1]{Ao Jin}
\ead{jinao@mail.nwpu.edu.cn}
\affiliation[1]{
	addressline={Research Center for Intelligent Robotics, School of Automation, Northwestern Polytechnical University, Xi\textquoteright an, China}}

\author[2]{Fan Zhang}
\ead{fzhang@nwpu.edu.cn}
\affiliation[2]{
	addressline={Research Center for Intelligent Robotics, School of Astronautics, Northwestern Polytechnical University, Xi\textquoteright an, China}}

\author[2]{Panfeng Huang\corref{mycorrespondingauthor}}
\ead{pfhuang@nwpu.edu.cn}

\cortext[mycorrespondingauthor]{Corresponding author at: Research Center for Intelligent Robotics, School of Astronautics, Northwestern Polytechnical University; E-mail address: pfhuang@nwpu.edu.cn}

%\credit{Conceptualization of this study, Methodology, Software}


\begin{abstract}
To avoid complex constraints of the traditional nonlinear method for tethered space robot (TSR) deployment, this paper proposes a data-driven optimal control framework with an improved deep learning based Koopman operator that could be applied to complex environments. In consideration of TSR's nonlinearity, its finite dimensional lifted representation is derived with the state-dependent only embedding functions in the Koopman framework. A deep learning approach is adopted to approximate the global linear representation of TSR. Deep neural networks (DNN) are developed to parameterize Koopman operator and its embedding functions. An auxiliary neural network is developed to encode the nonlinear control term of finite dimensional lifted system. In addition, the state matrix $A$ and control matrix $B$ of lifted linear system in the embedding space are also estimated during training DNN. Then three loss functions that related to reconstruction and prediction ability of network and controllability of lifted linear system are designed for training the entire network. With the global linear system produced from DNN, Linear Quadratic Regulator (LQR) is applied to derive the optimal control policy for the TSR deployment. Finally, simulation results verify the effectiveness of proposed framework and show that it could deploy tethered space robot more quickly with less swing of in-plane angle.
\end{abstract}

% Keywords

\maketitle

% Main text
\section{INTRODUCTION}

Satellite safety is gravely threatened by the abundance of space debris in the space environment. The space debris removal is extremely urgent for the space safety. There are many schemes to carry out the space debris removal missions, such as laser-based robot \cite{phipps2012clearing}, de-orbiting satellite using solar sails \cite{romagnoli2012orbiting}, flexible net \cite{bischof2003roger}, etc. The tethered space robot's advantages of strong flexibility, safe manipulation, low cost and transportability demonstrate its potential for space debris removal missions.

% the overview of control method for TSR
The control methods for tethered space robot (TSR) deployment have been studied a lot in the literature. So far, most control methods for TSR deployment, such as the backstepping control \cite{huang2016adaptive}, sliding mode control (SMC) \cite{Keshtkar2016,Li2022c,Li2020a}, discrete-time sliding model control (DSMC) \cite{Ma2020}, fuzzy control \cite{Xu2020} are nonlinear control methods. Other researchers have studied the linearizing the TSR's dynamics at the equilibrium point and then control the system with state feedback controller \cite{Pradeep1997}. Yet, stability and convergence are only maintained in the vicinity of the equilibrium point. As the initial state is far away from equilibrium point, system may be unstable under the controller designed by local linearization method. Some optimal control algorithms for TSR deployment have also been extensively studied \cite{Williams2008,Wen2008,Godard2010}.

It is promising to state that nonlinear control methods, such as SMC and DSMC for TSR deployment mentioned above, have many limitations. Most nonlinear controller are derived with some specific hypotheses, which may be not confirm with reality \cite{Hou2013}. The system parameters or model structure of some nonlinear plants may be time-varying, which brings great challenges for controller design. And some nonlinear control methods might be fail if the model is changed or disturbed. Many of challenges outlined above make it hard to design a stabilized controller for nonlinear system. Data-driven control (DDC) framework could alleviate these issues to a great extent, which merely utilizes the input and output data of the system to design controller. It's essential to propose a stronger control method in the DDC framework for TSR deployment mission.

Recently, the Koopman operator has demonstrated its advantages in dealing with nonlinear system, and some research results have been found in applying it for space applications \cite{Chen2020,servadio2023koopman}. The Koopman operator \cite{Koopman1931} was proposed by B.O.Koopman in 1931. It is a linear but infinite dimensional operator that governs the evolution of the embedding functions of nonlinear system's state. The basic idea of Koopman operator is to map a finite-dimensional nonlinear system to an infinite-dimensional linear system that is ready for implementation of linear control methods in an embedding space. And the finite-dimensional approximation of the Koopman operator gives a finite-dimensional linear system in the embedding space. It's noteworthy that the linearity of the Koopman operator is global, which is different from local linearization. With the state-dependent only embedding functions, one can derive the analytical finite-dimensional linear representation of a given nonlinear system \cite{2022arXiv220712132C}. But it's still a challenge to find the suitable embedding functions for a specific nonlinear system. 

\begin{figure}
	\includegraphics[width=18.5pc]{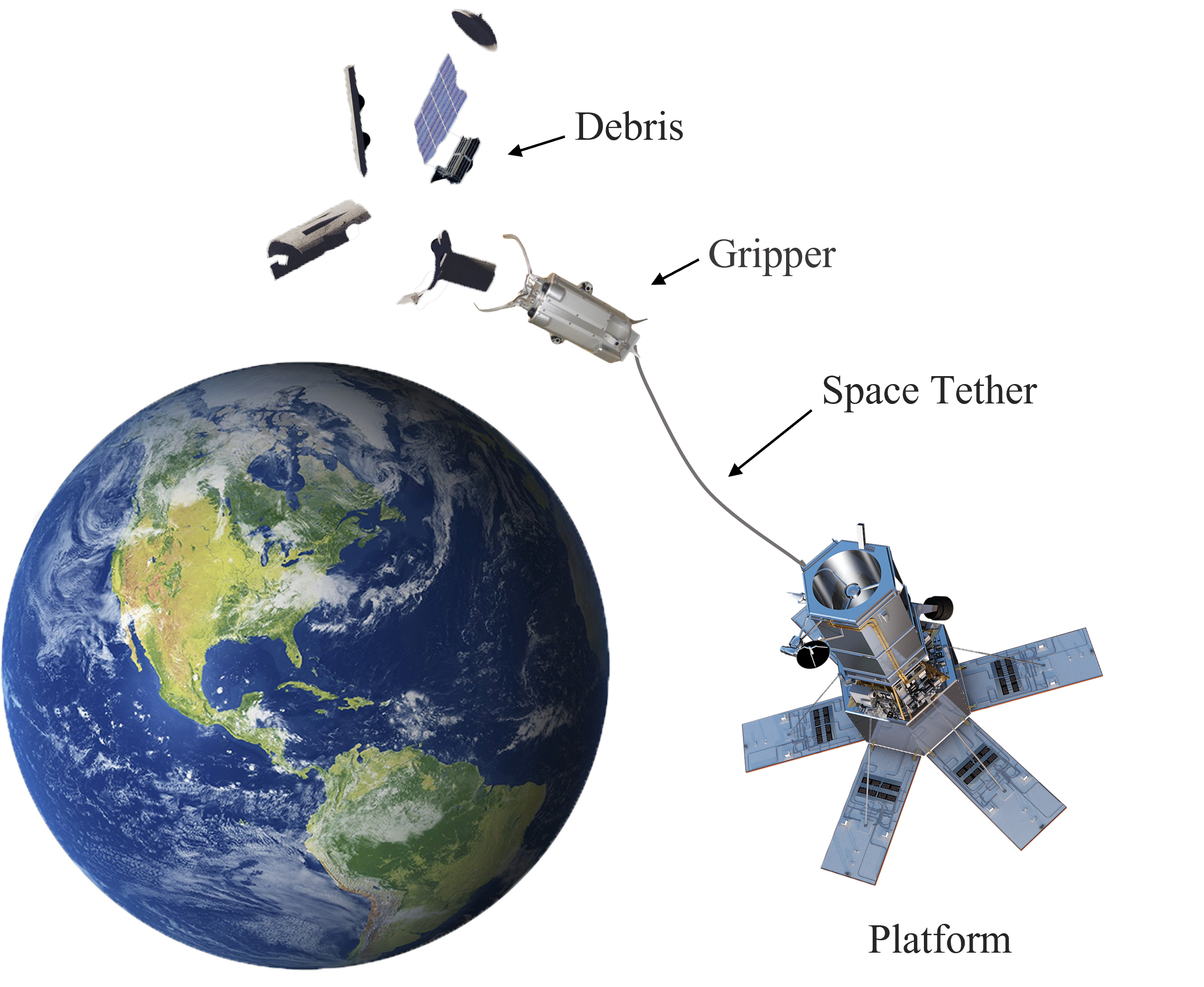} 
	\caption{Schematic diagram of tether space robot} \label{koopman}
\end{figure}

Some data-driven methods, such as the kernel method \cite{kevrekidis2016kernel}, dynamic mode decomposition (DMD) \cite{Proctor2016} and extending dynamic mode decomposition (EDMD) \cite{Williams2015}, approximate Koopman operator with pre-designed embedding functions, such as Hermite polynomials, radial basis functions and discontinuous spectral elements, for some simple nonlinear plants greatly. But in the case of sophisticated and high dimensional nonlinear system with control input, people found that these methods didn't work well for lack of generalization. As there is not a universal and effective framework to find the approximate embedding functions of a given nonlinear system especially the high dimensional nonlinear system until the deep learning scheme has shown its potential for finding embedding functions of complex nonlinear system \cite{Lusch2018} due to its strong ability of function approximation \cite{Liang2016}. And the literature \cite{Abraham2019} and \cite{Han2020} both mainly focus on the nonlinear system with control input using deep learning scheme and apply linear control methods to the plant. Recent deep learning based studies on Koopman operator can be found on \cite{Yeung2019,2020arXiv200609543H,Heijden2020,Shi2022,2022arXiv220812000W,Bakhtiaridoust2022,Wang2022c,2022arXiv221100357G,Zhao2023,gu2023deep}. The reconstruction and multiple steps prediction ability of network and controllability of lifted linear system in the embedding space were not taken into account together in these works.

Motivated by the points mentioned above, we proposed a data-driven optimal control framework for TSR deployment using the deep learning based Koopman operator in this work. The global linear representation of TSR dynamics with the embedding functions is derived firstly in the Koopman framework. Then we employ the proposed data-driven framework using deep neural network to find the embedding functions and approximate finite-dimensional Koopman operator with three designed loss functions which guarantee the reconstruction and prediction performance of the network and the controllability of lifted linear system. With the ability of prediction and the controllability, the lifted linear representation of TSR deployment is ready to apply linear control methods without complex nonlinear controller design. Additionally, analyzing the stability and convergence of a linear system is also easier than a nonlinear one, which can save a lot of efforts for controller designer. By combining the origin state with the output of lifting neural net together as the lifted state vector, the proposed data-driven framework does not require the decoder net. It could to recover the origin state from the output of network directly. In the stage of collecting data, the multi-threading technology is adopted to sample trajectories of TSR or other nonlinear systems, which can accelerate the data collection significantly on the modern hardware.

The main contributions of this work contain the following: 1) A data-driven optimal control framework with deep learning based Koopman operator is proposed for TSR deployment just using input and output data of plant, which could deploy TSR faster with less swing of in-plane angle. To the best of our knowledge, it is the first work that combines Koopman operator with optimal control for TSR deployment mission. 2) With the pre-defined state-dependent only embedding functions, a global and equivalent linear representation of TSR's nonlinear dynamics is derived in the deep learning based Koopman operator framework, which could integrate with linear control methods directly. 3) An enhanced deep learning method is proposed for finding embedding functions and finite-dimensional approximation of Koopman operator without decode state from embedding space, and three specific defined loss functions are raised to guarantee reconstruction and multiple steps prediction ability of network and controllability of lifted linear system in the embedding space.

This paper is organized as follows. Section I reviews the recent developments on nonlinear control methods with deep learning based Koopman operator and control methods for TSR deployment. Section II gives a brief presentation 
on the dynamics of TSR deployment for collecting data and the basis of Koopman operator for nonlinear system with control input. Section III introduces the proposed data-driven optimal control framework with deep learning based Koopman operator for TSR deployment. Section IV illustrates the details of network training and presents the prediction and control performance of proposed method compared with other methods. Section V concludes this paper.

\section{System Dynamics and Koopman Operator}

\subsection{Dynamics of Tether Space Robot}
The dynamics of TSR for deployment derived in \cite{Williams2008} is excited to collect data for training in this work. For sake of simplicity, the dimensionless ordinary differential equation (ODE) of TSR is given directly as

\begin{gather}
	\ddot{\theta}=2(\dot{\theta}+1)[\beta \tan{\beta}-\frac{\dot l}{l}]-3\sin{\theta}\cos{\theta} \notag
	\\
	\ddot{\beta} = -2\frac{\dot l}{l}\dot{\beta} - [(\dot{\theta}+1)^2+3{\cos}^2{\theta}]\sin{\beta}\cos{\beta} \notag
	\\
	\ddot{l} = l[{\dot{\beta}}^2+(\dot{\theta}+1)^2{\cos}^2\beta+3{\cos}^2\theta{\cos}^2\beta -1] - u
\end{gather}
and 
\begin{equation}
	l=x/L,u=T/(m{\Omega}^2L),\tau=\Omega t \notag
\end{equation}
where $x$ and $L$ are the instantaneous and total length of space tether respectively, $l\in (0,1]$ is the tether deployment ratio, $T$ is the tension of space tether, $m$ is the mass of TSR, $\theta$ is the in-plane angle of TSR, $\beta$ is out-plane angle, $\Omega$ represents the orbit angular velocity, and $\tau$ is the dimensionless form of time $t$. For the in-plane deployment ($\beta = \dot{\beta}=0$) of TSR, the dynamics can be written as
\begin{equation}
	\dot{{X}} = {f}({X}) + {b}u
	\label{state_dynamics}
\end{equation}
where $x_1=\theta,x_2=\dot{\theta},x_3=l-1,x_4=\dot{l},{X}=[x_1,x_2,x_3,x_4]$, with
\begin{gather}
	{f}({X})=
	\left[ \begin{array}{c}
		x_2\\
		-2\frac{x_4}{x_3+1}(x_2+1)-3\sin x_1\cos x_1\\
		x_4\\
		\left( x_3+1 \right) \left[ \left( x_2+1 \right) ^2+3\cos ^2x_1-1 \right]\\
	\end{array} \right] \notag
	\\
	{b}=[0,0,0,-1]^T	\notag
	\\
	-1 \textless x_3 \le 0,0 \le u 
\end{gather}
The equilibrium point of (\ref{state_dynamics}) is $[x_1,x_2,x_3,x_4] = [0,0,0,0]$, which is also the final state of TSR deployment. With the assumption that tether is straight during deployment, it is required that the tension of tether need to be positive i.e. $u\ge 0$. The control problem of TSR deployment is to drive TSR from its initial state $X_0 = [0,0,l_0-1,v_0]$ to the final state $X_f = [0,0,0,0]$. It's worth noting that when the TSR reaches the final state, there is a constant and positive control input $u=3$ acting on the space tether. Equation (\ref{state_dynamics}) is used to represent the dynamics of TSR in the latter part of this paper.

\subsection{Preliminaries on Koopman Operator}
Consider a discrete-time nonlinear system without control input in the general form
\begin{equation}
	x_{k+1}=f(x_k)  
\end{equation}
where $x_k\in \mathbb{R}^n, k \in \mathbb{N}$.

We define a vector-valued observable ${\Phi (x_k)}$ which is the function of state. Each function in ${\Phi}$ is an element of an infinite dimensional Hilbert space $\mathcal{H}$. The Koopman operator, an infinite linear operator, acts on the observable
\begin{equation}
	\mathcal{K} {\Phi} = {\Phi} \circ f
\end{equation}
where $\circ$ is the composition operator, so that
\begin{equation}
	\mathcal{K} {\Phi}(x_k) = {\Phi}(x_{k+1})
\end{equation}

Namely, the Koopman operator $\mathcal{K}$ defines a linear but infinite dimensional system that advances the state-dependent observable $\Phi(x_k)$ to the next timestep. It should be noted that Koopman operator acts on the embedding functions or the observable ${\Phi}$, rather than the state $x$. Therefore, by lifting the origin state space to an infinite dimensional and invariant Hilbert space $\mathcal{H}$ with observable ${\Phi}$, the Koopman operator allows one to convert a nonlinear system into a linear system. In this way, we could deem that the Koopman operator trades linearity with dimension. More details about Koopman operator, readers can refer to \cite{Proctor2018,Brunton2016,Bevanda2021,Otto2021,Bruder2021a}.

Now consider the general nonlinear system with control input
\begin{equation}
	x_{k+1} = f(x_k,u_k) \label{koopman_control}
\end{equation}
and split the nonlinear function $f(x_k,u_k)$ in (\ref{koopman_control}) into two parts
\begin{equation}
	x_{k+1} = f(x_k,u_k) = f(x_k,0) + g(x_k,u_k) \label{koopman_sep}
\end{equation}
where $x_k\in \mathbb{R}^n, u_k\in \mathbb{R}^m, k \in \mathbb{N}$. The first item $f(x_k,0)$ is called autonomous dynamics by setting $u_k=0$, and the second item $g(x_k,u_k)$ is called control driven dynamics. By applying state-dependent only observable ${\Phi}$ to (\ref{koopman_sep}), a lifted linear system can be derived. 

\begin{theorem}\cite{2022arXiv220712132C}  \label{thm1}
	Given a nonlinear system (\ref{koopman}) and a state-dependent only observable ${\Phi}$ where ${\Phi} (f(x,0)) \in span \{{\Phi}\}$ with ${\Phi}: X \rightarrow \mathbb{R}^K$ and $X$ is convex, then an exact finite dimensional lifted form of the nonlinear system is defined as:
	\begin{equation}
		{\Phi}(x_{k+1}) = A{\Phi}(x_k)+\mathcal{B}(x_k,u_k) \label{theme1}
	\end{equation}
	with $A\in \mathbb{R}^{K \times K} $ and
	\begin{align}
		&\mathcal{B}(x_k,u_k) = \notag
		\\
		&\left(\int_{0}^{1}{\frac{\partial\Phi}{\partial x}(f(x_k,0)+\lambda g(x_k,u_k))}d\lambda \right)g(x_k,u_k)
	\end{align}
\end{theorem} 
For the nonlinear dynamics with linear control input form 
\begin{equation}
	x_{k+1} = f(x_k)+bu_k \label{linear_control}
\end{equation}
the application of Theorem \ref{thm1} leads to
\begin{equation}
	\mathcal{B} (x_k,u_k)=\underset{B\left( x_k,u_k \right)}{\underbrace{\left( \int_0^1{\frac{\partial {\Phi}}{\partial x}(f(x_k)+\lambda bu_k)}d\lambda \right) }}bu_k
\end{equation}
Now by employing the pre-defined observable ${\Phi}$,  the nonlinear system (\ref{linear_control}) with linear control input has a lifted form
\begin{equation}
	{\Phi}(x_{k+1}) = A{\Phi}(x_k)+B(x_k,u_k)u_k \label{TSR_linear}
\end{equation}

The dynamics of TSR has the same form as (\ref{linear_control}). A finite dimensional representation of TSR can be deduced directly with the pre-defined state-dependent only observable $\Phi$ by employing Theorem \ref{thm1} in the Koopman framework. It's worth noting that system (\ref{TSR_linear}) is linear in state $x$ but not linear in control input $u$, which cannot integrate with linear control scheme directly. In the next section, the proposed framework would convert (\ref{TSR_linear}) into a completely linear form.

\begin{figure}
	\includegraphics[width=18.5pc]{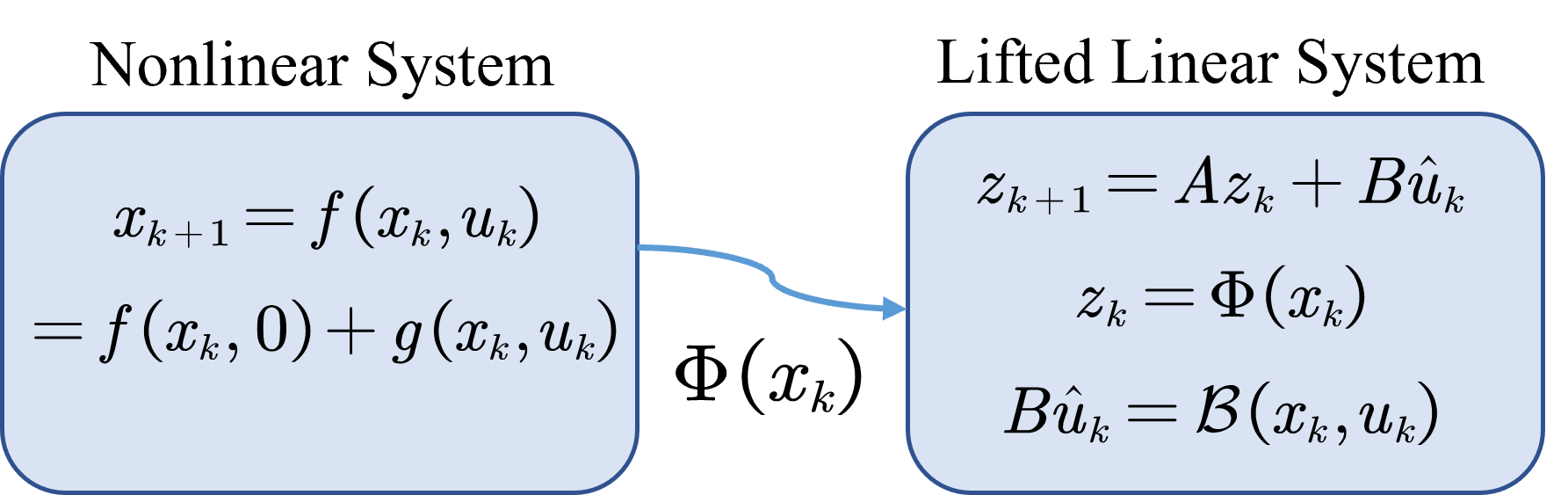} 
	\caption{Illustration of Koopman operator for nonlinear system with control input} \label{koopman}
\end{figure}

\section{Koopman Operator Learning and Controller Design}
In this section, an enhanced data driven method is proposed to approximate the global and equivalent linear representation of TSR's dynamics using deep learning based Koopman operator. Given the importance of resource consumption during TSR deployment missions, a optimal control policy based on Linear Quadratic Regulator is applied to the lifted linear system obtained from the deep neural network. 

\subsection{A Data Driven Framework for learning Koopman Operator}
To linearize the nonlinear dynamics in the Koopman operator framework, the first step is to find the observable or embedding functions that could map the state of the nonlinear system to a higher dimensional invariant space $\mathcal{H}$. In this work, we are going to utilize a data-driven framework with deep neural network to find the observable and approximation of Koopman operator. 

\begin{figure}
	\includegraphics[width=18.5pc]{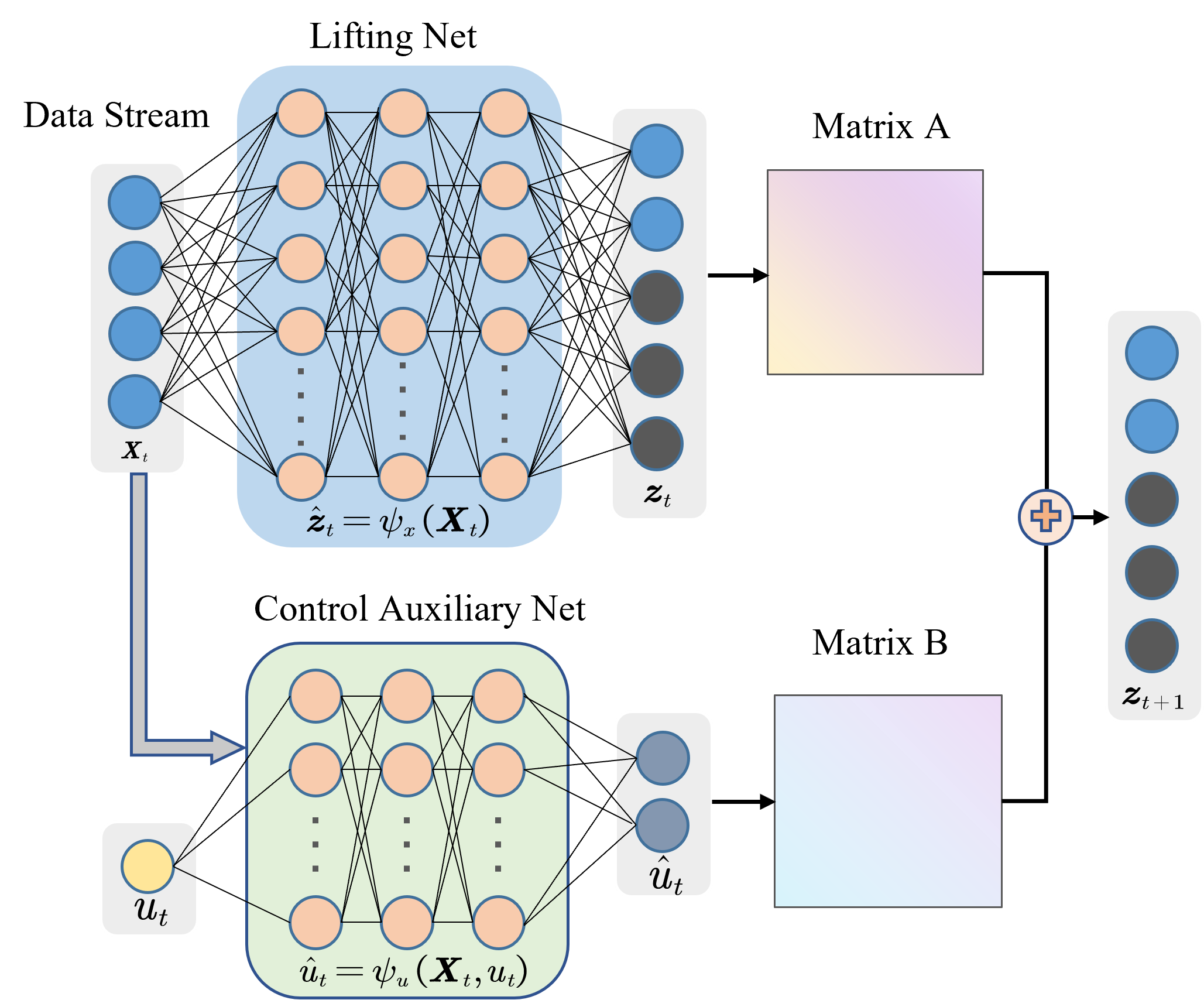} 
	\caption{Framework for learning Koopman operator.} \label{fig1}
\end{figure}

The data-driven framework for finding the embedding functions and learning Koopman operator is shown in Fig.\ref{fig1}. The embedding function $\Phi(x_k)$ in (\ref{TSR_linear}) is parameterized with deep neural network, resulting to $\Phi_{\theta}(x_k)$. The $\theta$ represents the parameters of deep neural networks. An embedding neural network $\psi_{x}(x_k)$ mapping $x_k$ to a higher-dimensional space $\mathcal H$ and a control auxiliary net $\psi_u (x_k,u_k)$ are also developed for embedding control input. The lifted linear system (\ref{theme1}) is derived from applying state-dependent only observable $\Phi$ to the nonlinear system (\ref{koopman}) with control input. The term $\mathcal{B}$ in (\ref{theme1}) still remains nonlinear, which resulting that it's hard to capture the full inherent dynamics of complex nonlinear system from data. So we decompose the term $B(x_k,u_k)u_k$ of (\ref{TSR_linear}) into two parts $B$ and $\psi_u(x_k,u_k)$. Thus, we can rewrite (\ref{TSR_linear}) in a purely linear form 

\begin{equation}
	{z}_{k+1} = A{z}_{k}+B\hat{u}_k \label{lifting_dynamics}
\end{equation}
where
\begin{gather}
	{z}_k = \psi_{x}(x_k) \triangleq
	\left[  \begin{array}{c}
		x_k\\
		\Phi_{\theta} \left( x_k \right)\\
	\end{array}  \right]  \label{state_lifting}  
	\\
	\hat{u}_k = \psi_u (x_k,u_k) \label{control_lifting}
\end{gather}
and $A \in \mathbb{R}^{({K+n})({K+n}}),B\in \mathbb{R}^{(K+n)p},x_k \in \mathbb{R}^n,{z}_k \in \mathbb{R}^{K+n},\hat{u}_k \in \mathbb{R}^p,  \Phi_{\theta}:\mathbb{R}^n \rightarrow \mathbb{R}^K, K \gg n$. The state matrix $A$ and control matrix $B$ are estimated during training the deep neural network.

In contrast to \cite{Lusch2018}, the data-driven framework that we proposed joint the origin state $x_k$ and $\Phi_{\theta}(x_k)$ together as the state vector of lifted linear system. Thus it does not require another network to decode origin system state from embedding space $\mathcal{H}$ and it could recovers origin state from embedding space straightforwardly
\begin{equation}
	\hat{x}_k = C{z}_k \label{x_recover}
\end{equation}
where $C=
\left[ \begin{matrix}
	I_n&		{0}_K\\
\end{matrix} \right] 
$. Combining (\ref{state_lifting}) with (\ref{x_recover})
\begin{equation}
	\hat{x}_k =  C\psi_{x}(x_k) \label{recover}
\end{equation}
%where $\hat{x}$ is defined as output of network $\psi_{\theta}$.

Using (\ref{recover}), it's easy to recover origin state $\hat{x}$ from embedding space. As there is not a decoder neural network, the proposed network has fewer parameters and a more concise structure than the network based on the deep autoencoder \cite{Lusch2018,Azencot2020,Masti2021,Zhao2023}, which means it takes less time to train the network on even ground.

\subsection{Network Performance Index}
The architecture of proposed network are depicts in the Fig.\ref{fig1}. The objective of this network is to identify a global linear and equivalent expression of (\ref{state_dynamics}). To guarantee the reconstruction and prediction ability of network and stability of lifted linear system, three loss functions that characterize the performance of network are developed as follows. 

\begin{enumerate}
	\item 
	\textbf{Loss function of state reconstruction}\\
	Linear system (\ref{lifting_dynamics}) evolves under a higher dimensional space $\mathcal{H}$. The origin state of a nonlinear system could be reconstructed using the network $\psi_{x}$. The reconstruction accuracy of network $\psi_{x}$ is evaluated by 
	\begin{equation}
		L_1(x_k) = \Vert x_k-\hat{x}_k \Vert = \Vert x_k-C\psi_{x}(x_k) \Vert \label{L1}
	\end{equation}
	\item 
	\textbf{Loss function of multi-steps prediction}\\
	By iterating (\ref{lifting_dynamics}) with a given initial state $z_0$ and control input, one can easily predict the state of lifted system after $m$ steps. Given a dataset
	\begin{equation}
		\left[X_i\in \mathbb{R}^{n\times m},U_i \in \mathbb{R}^{p\times m},i=1,...,k \right] \notag
	\end{equation}
	contains $k$ trajectories and each trajectory $[X_i,U_i]$ contains $m$ steps data shown in Fig.\ref{fig4}. The actual state of lifted system could be computed with $Z_i=\psi_x(X_i), Z_i\in \mathbb{R}^{(K+n)\times m}$ and predicted state by iterating (\ref{lifting_dynamics}) is denoted $\hat Z_i$. The prediction error of a dataset contains $k$ trajectories is defined as 
	\begin{equation}
		L_2(\psi_{x}) = \sum_{i=1}^{k} \gamma^{i} {\rm MSE}(Z_{i},\hat{Z}_{i}) \label{L2}
	\end{equation}
	where $\gamma \in(0,1)$ is a decay hyperparameter and MSE denotes the mean square error evaluating difference between the estimated value and the actual value. Unlike one step state prediction index, loss function (\ref{L2}) focuses the sum of multiple steps prediction error, which is helpful for obtaining more precise prediction over a longer horizon.
	\item 
	\textbf{Controllability of lifted linear system}\\
	By applying the deep network shown in Fig.\ref{fig1}, the TSR's nonlinear dynamics is transformed into a linear system (\ref{lifting_dynamics}). While it might be uncontrollable in the embedding space $\mathcal{H}$. If the lifted linear system is controllable, the stability and convergence of (\ref{lifting_dynamics}) can be guaranteed. The necessary and sufficient condition for controllability of (\ref{lifting_dynamics}) is $rank(B,AB,...,A^{K-1}B) = K$. So we define a loss function to guarantee the controllability of the lifted linear system. 
	\begin{align}
		L_3&(A,B) = \notag
		\\	
		&\Vert K-rank(B,AB,...,A^{K-1}B) \Vert  \label{L3}
	\end{align}
\end{enumerate}

Then, combining (\ref{L1}), (\ref{L2}) and (\ref{L3}), we give the loss function of the entire network shown in Fig.\ref{fig1} 
\begin{equation}
	L = \alpha L_1 + \beta L_2 + \eta L_3 \label{loss_all} 
\end{equation}
where $\alpha$, $\beta$ and $\eta$ are three user-defined hyperparameters. In the actual training stage, we usually choose a smaller $\eta$ than $\alpha$ and $\beta$, since $L_3$ is always decreasing more quickly than $L_1$ and $L_2$. More details about training will be discussed in the simulation section later. 

\begin{figure}
	\centering
	\includegraphics[width=18.5pc]{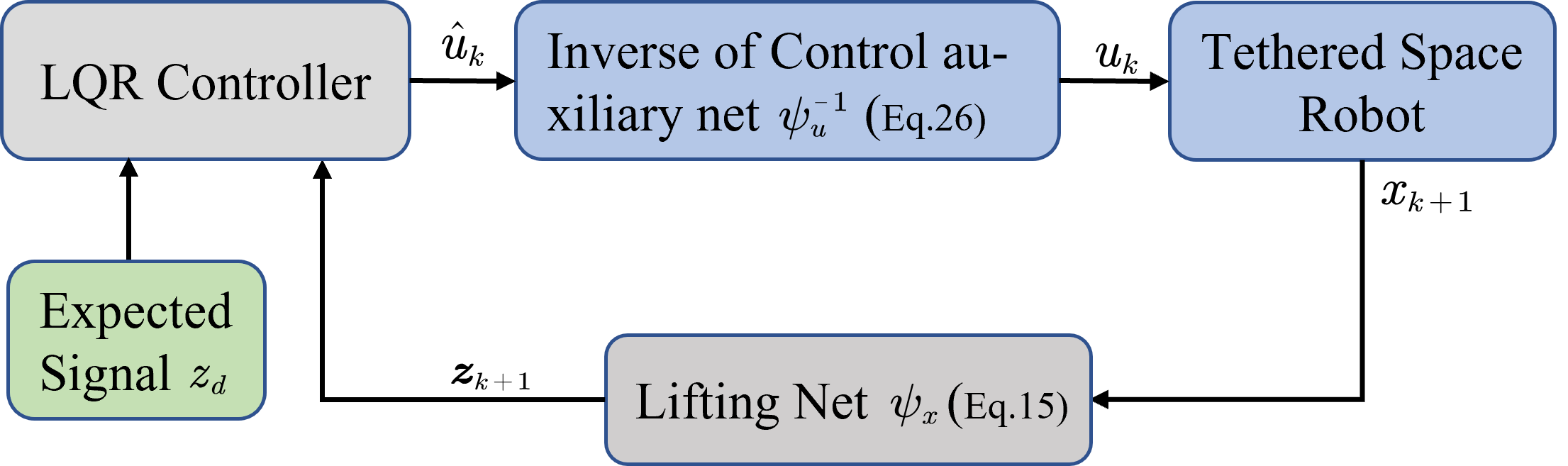} \label{fig3}
	\caption{Schematic of optimal control for TSR deployment using deep learning based Koopman operator}
\end{figure}

\subsection{Linear Quadratic Regulator}
Given the initial state $x_0$ and expected state $x_d$ and expected control $u_d$ of the nonlinear system (\ref{koopman_control}), the optimal control problem with quadratic performance index and infinite horizons can be represented as
\begin{gather}
	J=\sum_{i=0}^{\infty}\left[({x_i-x_d})^TQ({x_i-x_d})+({u_i-u_d})^TR(u_i-u_d)   \right] \notag
	\\
	\mbox{s.t.}\quad x_{k+1}=f(x_k,u_k) \label{nonlinear_opti}
\end{gather}
where $Q$ and $R$ are the weight matrix related to state and control respectively.

As we know, nonlinear optimal control problems are hard to solve directly. Fortunately, the Koopman operator could be capable of projecting a nonlinear system (\ref{koopman}) onto a corresponding linear system (\ref{lifting_dynamics}) globally. Therefore, many linear control methods could be applied to the nonlinear system. With (\ref{lifting_dynamics}) and (\ref{recover}), we rewrite the optimal control problem (\ref{nonlinear_opti}) for the nonlinear system as
\begin{gather}
	J=\sum_{i=0}^{\infty}\left[({{z}_i-  {z}_d})^TQ'({{z}_i-{z}_d})+({\hat{u}_i-\hat{u}_d})^TR'(\hat{u}_i-\hat{u}_d)   \right] \notag
	\\
	\mbox{s.t.}\quad {z}_{k+1} = A{z}_k+B\hat{u}_k \notag
	\\
	{z}_0 = \psi_{x}(x_0),{z}_d = \psi_{x}(x_d)  \label{lqr_index} \notag
	\\
	\hat{u}_k = \psi_u(x_k,u_k),\hat{u}_d = \psi_u(x_d,u_d) \label{lqr}
\end{gather} 
where $Q'=C^TQC$ and $R'=R$. Using dynamic programming to minimize the index (\ref{lqr}), the optimal feedback controller gain $K_{LQR}$ of (\ref{lifting_dynamics}) could be obtained. The control law is
\begin{equation}
	\hat{u}^*_k-\hat{u}_d = K_{LQR}({z}_k-{z}_d)
\end{equation}

Recover the real control input that will be applied to TSR by using (\ref{control_lifting})
\begin{equation}
	u_k^* = {\psi}^{-1}_u(x_k,\hat{u}^*_k) \label{opti_control}
\end{equation}

The detailed implementation of the LQR control based on Koopman operator is illustrated in Algorithm \ref{algorithm1}. For simplicity, we assumed that the term $\psi_u(x_k,u_k)$ of (\ref{control_lifting}) is the product of $\psi_{ux}(x_k)$ and $\psi_{uu}(u_k)$, avoiding to compute inverse of network $\psi_u(x_k,u_k)$, which is typically a complex job. So (\ref{opti_control}) can be rewritten as
\begin{equation}
	u^*_k = \frac{\psi_{ux}(x_k)}{\psi_{uu}(\hat{u}^*_k)} \label{opti_real_control}
\end{equation}

\begin{figure}[!t]
	\begin{algorithm}[H]
		\caption{LQR control for TSR based Koopman operator}
		\label{algorithm1}
		\begin{algorithmic}[1]
			\REQUIRE TSR dynamics $f(x)$, initial and expected state $x_0$, $x_d$, lifting net $\psi_{x}$, control net $\psi_u$, matrix A and matrix B, LQR weighed matrix $Q, R$
			\STATE Initialization: initial the LQR gain $K_{LQR}$ by dynamic programming, ${z}_0 \leftarrow \psi_{x}(x_0)$, ${z}_0 \leftarrow \psi_{x}(x_d)$
			\WHILE {TSR state $x_k$ does not reach expected state $x_d$}
			\STATE lift the state ${z}_k \leftarrow \psi_{x}(x_k)$
			\STATE compute controller output $\hat{u}^*_k \leftarrow L_{LQR}({z}_k-{z}_d)$
			\STATE recover real controller output $u^*_k \leftarrow \psi^{-1}_u(x_k,\hat{u}^*_k)$
			\STATE apply $u^*_k$ to TSR $x_{k+1} = f(x_k,u_k)$
			\STATE $k = k+1$
			\ENDWHILE
		\end{algorithmic} 
	\end{algorithm} 
\end{figure}

\section{Simulation and Results}
In this section, we demonstrate the effectiveness and superiority of the data-driven framework we proposed for the TSR in-plane deployment. We assume that the model of TSR is unknown and only the input/output data is available for the model identification in the Koopman framework and controller design. And the detailed implementation of the network training will also be illustrated in this section.

\begin{table}[!t]
	\renewcommand{\arraystretch}{1.3}
	\caption{Range of initial state for sampling}
	\centering
	\label{table_1}
	%\centering
	\resizebox{\columnwidth}{!}{
		\begin{tabular}{ccc}
			\hline\hline \\[-3mm]
			\multicolumn{1}{c}{Variables} & \multicolumn{1}{c}{Value Range}   & \multicolumn{1}{c}{Distribution} \\[1.2ex] \hline
			$x_1$ & [-2.5,0.5] & uniform distribution \\
			$x_2$ & [-1,1] & uniform distribution \\
			$x_3$ & [-0.99,0] & uniform distribution \\
			$x_4$ & [0,2] & uniform distribution \\ 
			$u$ & [0,5] & uniform distribution \\
			
			\hline\hline
		\end{tabular}
	}
\end{table}

\subsection{Data Sampling}
\begin{figure}[!t]
	\centering
	\includegraphics[width=18.5pc]{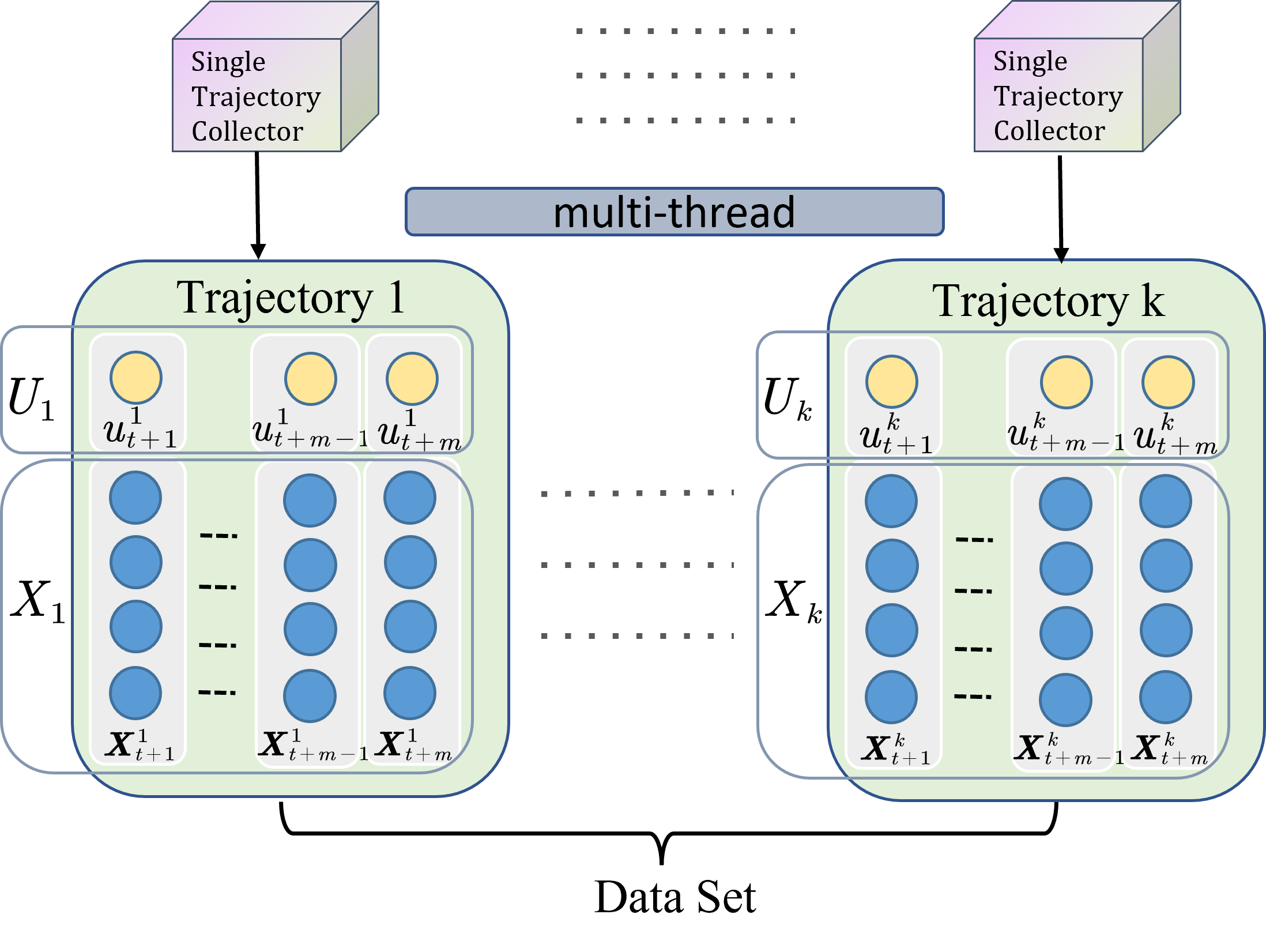} 
	\caption{The framework of data sampling using multi-thread} \label{fig4}
\end{figure}

To obtain dataset, the dynamics of TSR (\ref{state_dynamics}) is discretized using Runge–Kutta four method with the sampling period $T_s = 0.01 {\rm s}$ and excited with random control input. And the initial conditions of each trajectory in the dataset are generated randomly with a uniform distribution. The range of initial state $x_1,x_2,x_3,x_4$ and control input $u$ can be found in Table.\ref{table_1}. In the stage of sampling dataset, the multi-threading technology is adopted to collect data from the plant or simulator. Sampling the same-sized dataset, multi-thread technology is much quicker compared to single-thread technology on the modern hardware, especially in case of sampling big scale dataset. The details of sampling in this paper are shown in Fig.\ref{fig4}.

\subsection{Network Parameters Setting}
In our experiments, the lifting network and the control auxiliary network both have four hidden layers and each layer has 128 units, the matrix $A$ and matrix $B$ only have one layer with $K+n$ units. The hyperparameters of the loss function (\ref{loss_all}) are set $\gamma = 1-10^{-3}, \alpha = 0.5, \beta = 1, \eta = 0.1$.

\subsection{Accuracy of reconstruction and prediction}
We train the network with parameters mentioned above, varying in $[1,2,3,4,5,6]\times 10^4$ trajectories, and each trajectory contains 30 steps data of TSR's dynamincs. The dimension of lifting space is chosen as $K=36$. As the size of dataset is very large, the mini-batch optimization \cite{Li2014} is adopted in the training stage. The prediction error of a dataset in this section is evaluated by
\begin{gather}
	Error = \sum_{i=1}^{k} {\rm MSE}(X_i,\hat{X}_i) = \sum_{i=1}^{k} {\rm MSE}(X_i,C\psi_x(X_i)) \label{prediction_error}
\end{gather}
which focuses on the reconstruction and prediction error of the lifted system. 

\begin{figure}[!t]
	\centering
	\includegraphics[width=18.5pc]{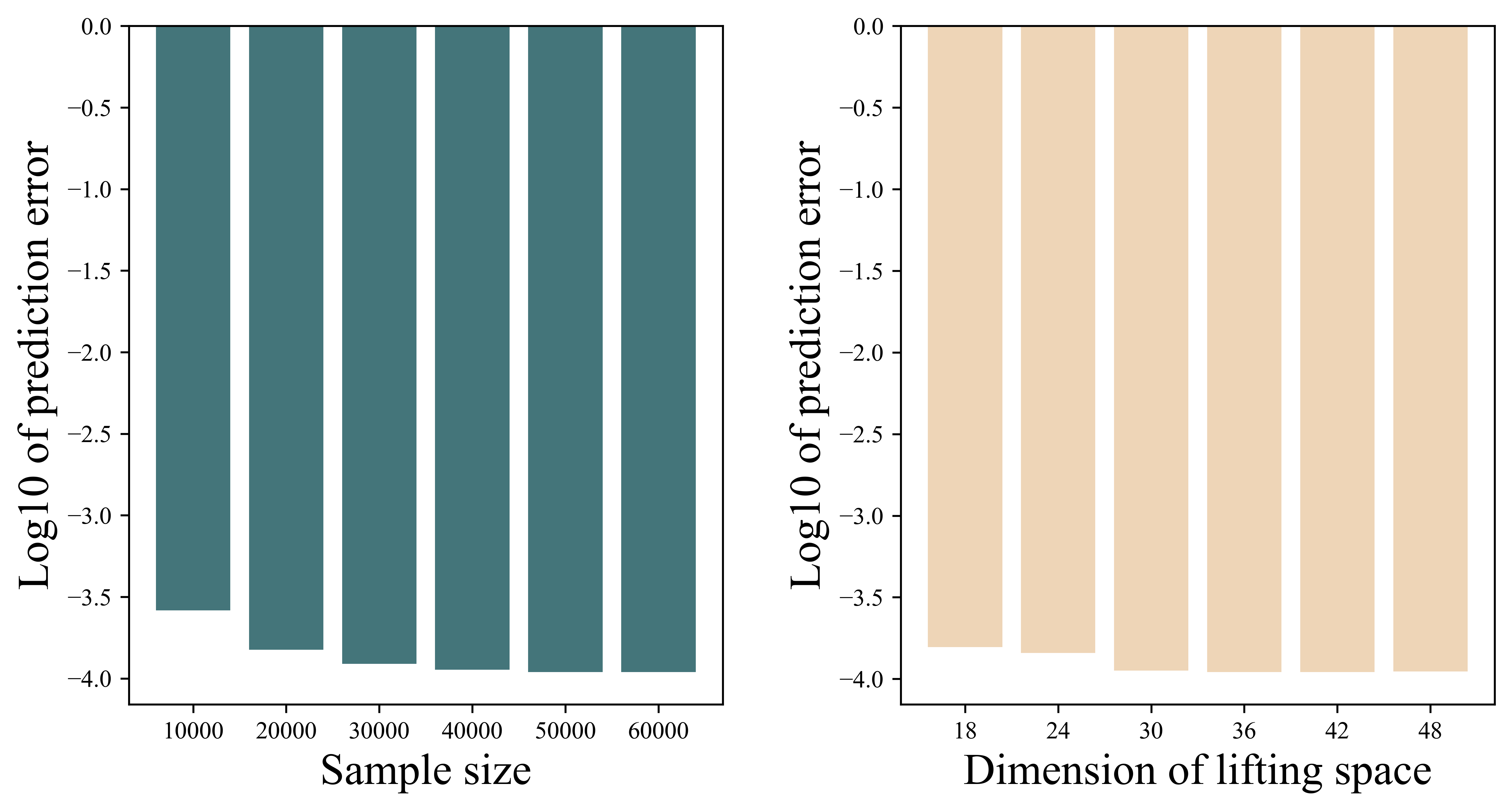} 
	\caption{Prediction error with different sample size and lifting dimension} \label{samplesize}
\end{figure}

The prediction performance of the proposed framework is depicted in the left panel of Fig.\ref{samplesize}. It's interesting that the prediction error is not descending when the dataset contains over 40000 trajectories. The invariant space of lifted linear system may be captured by the deep neural network totally. And the log of the prediction error is around $1 \times 10^{-4}$. By statements of the foregoing, the proposed framework could be capable of learning a good linear approximation of TSR dynamics just requiring 40 thousand trajectories. 

Additionally, the impact of lifting space dimension $K$ on the prediction accuracy of linear lifted system is explored. The size of the training dataset is 40000 in this simulation. The result is plotted in the right of Fig.\ref{samplesize}. It's worth that when $K>36$, prediction error does not decrease with increasing $K$. This contradicts the conclusion of Theorem.2 in \cite{Korda2018}, which indicates that the approximation of Koopman operator would converge to the true one as the $K$ goes infinity. With the embedding functions found by deep network, the lifted dynamics in the embedding space is nearly linear as $K=36$. And the dimension of true Koopman operator of TSR's dynamics might be 36 according the results. Thus when the $K>36$, the prediction error is not descending. And the invariant Hilbert space $\mathcal{H}$ in the Koopman framework is finite dimensional in the case of TSR's dynamics in particularly. 

\begin{figure}[!t]
	\centering
	\includegraphics[width=18.5pc]{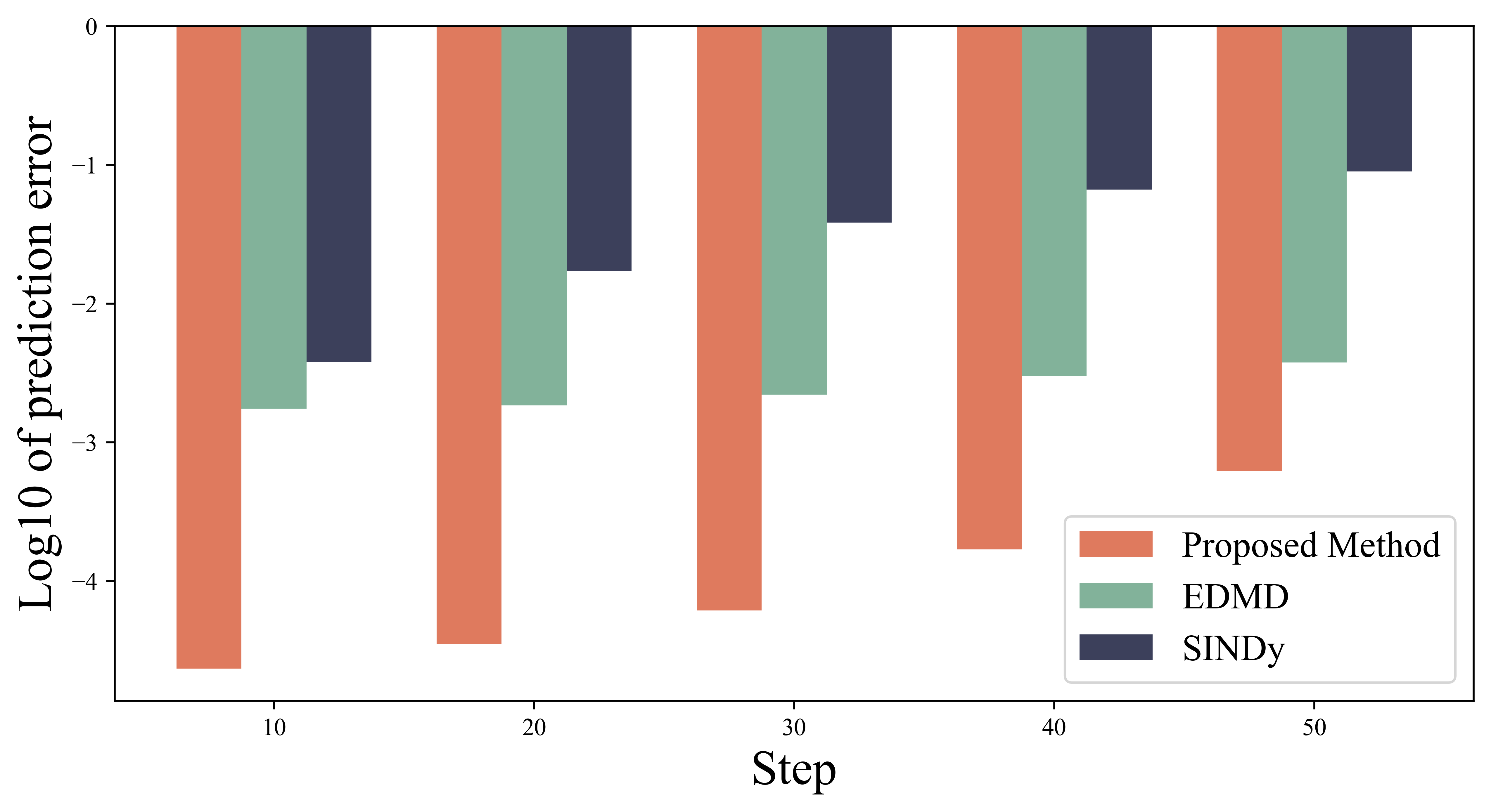} 
	\caption{Prediction error of different methods} \label{methods}
\end{figure}

The prediction performance of proposed deep learning framework is compared with EDMD \cite{Korda2018} and SINDy \cite{Brunton2016a}. The size of training dataset is 40000 and dimension of lifted linear system is $K=36$ in the comparison simulation. The lifting functions of EDMD are chosen to be thin spline radial basis functions. Four-order polynomial library is chosen as the library function of SINDy. The result is plotted in the Fig.\ref{methods}. Although prediction error decreases with prediction step increases among three methods, the proposed deep learning framework has a better prediction performance over a long time horizon than EDMD and SINDy. And the prediction error of proposed deep learning framework is much smaller, which indicates that the proposed framework could capture the inherent dynamics of nonlinear system entirely. In the case of TSR's dynamics, the proposed deep learning framework owns superior prediction performance than EDMD and SINDy. 

The prediction results mentioned above indicate that the embedding functions and Koopman operator can be approximated greatly by the proposed deep neural network, which results a global linear representation of TSR dynamics.

\subsection{Control Simulation for TSR deployment}
As for control performance, we compared our method with \cite{Sun2014} and \cite{Pradeep1997}, which have wide application in the TSR deployment missions. To be fair, the initial states of TSR deployment are assumed to be $x_1=0,x_2=0,x_3=-0.99,x_4=0.5$ ,which is the same as that in \cite{Sun2014} and \cite{Pradeep1997}. And the physical parameters, including orbit rate $\Omega=1.1804\times 10^{-3} {\rm rad}/s$ and tether length $L=100{\rm km}$, are also identical to those of \cite{Sun2014}.

\begin{figure}[!t]
	\centering
	\includegraphics[width=18.5pc]{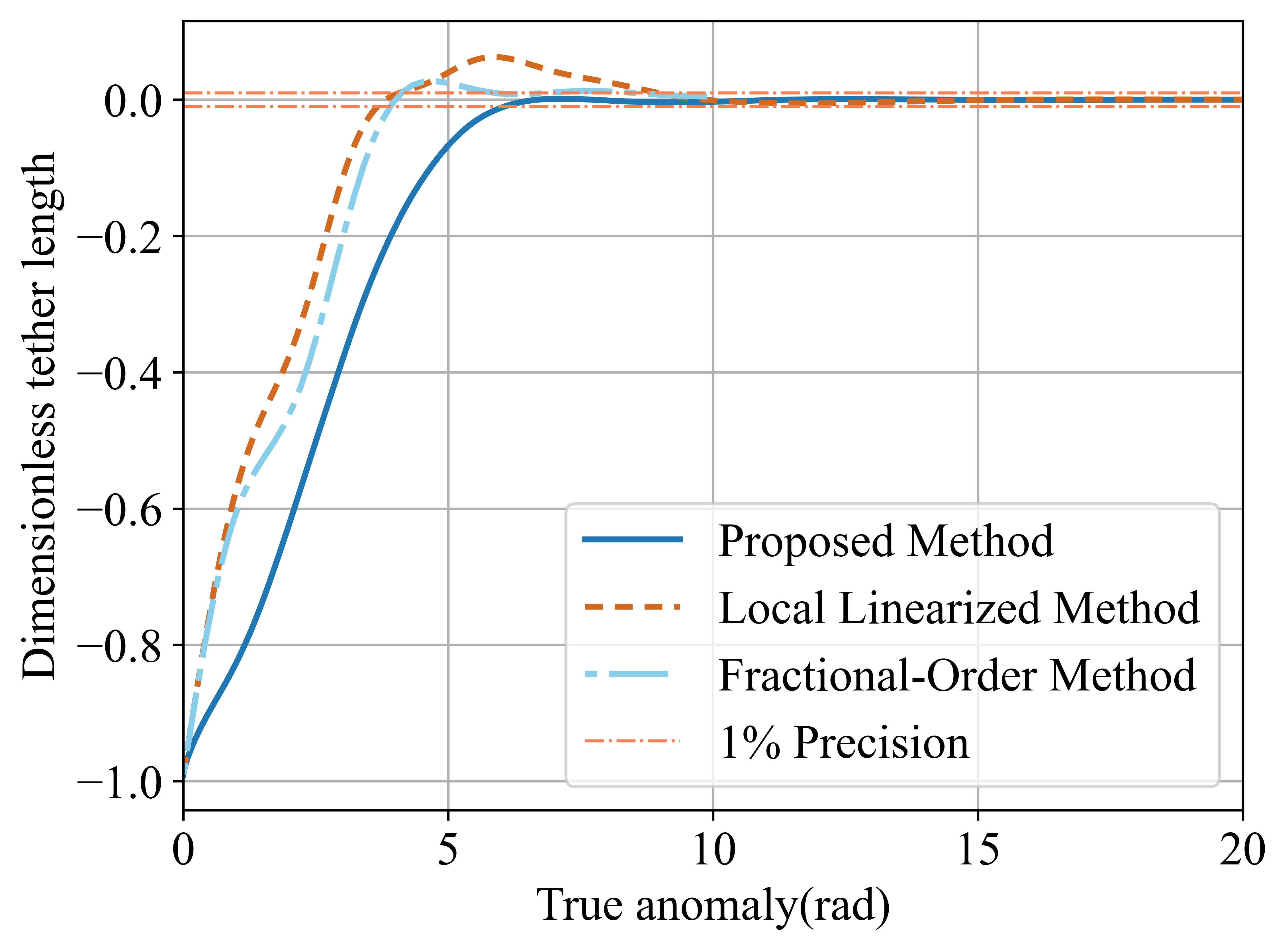} 
	\caption{Trajectory of dimensionless tether length} \label{fig5}
\end{figure}

\begin{figure}[!t]
	\centering
	\includegraphics[width=18.5pc]{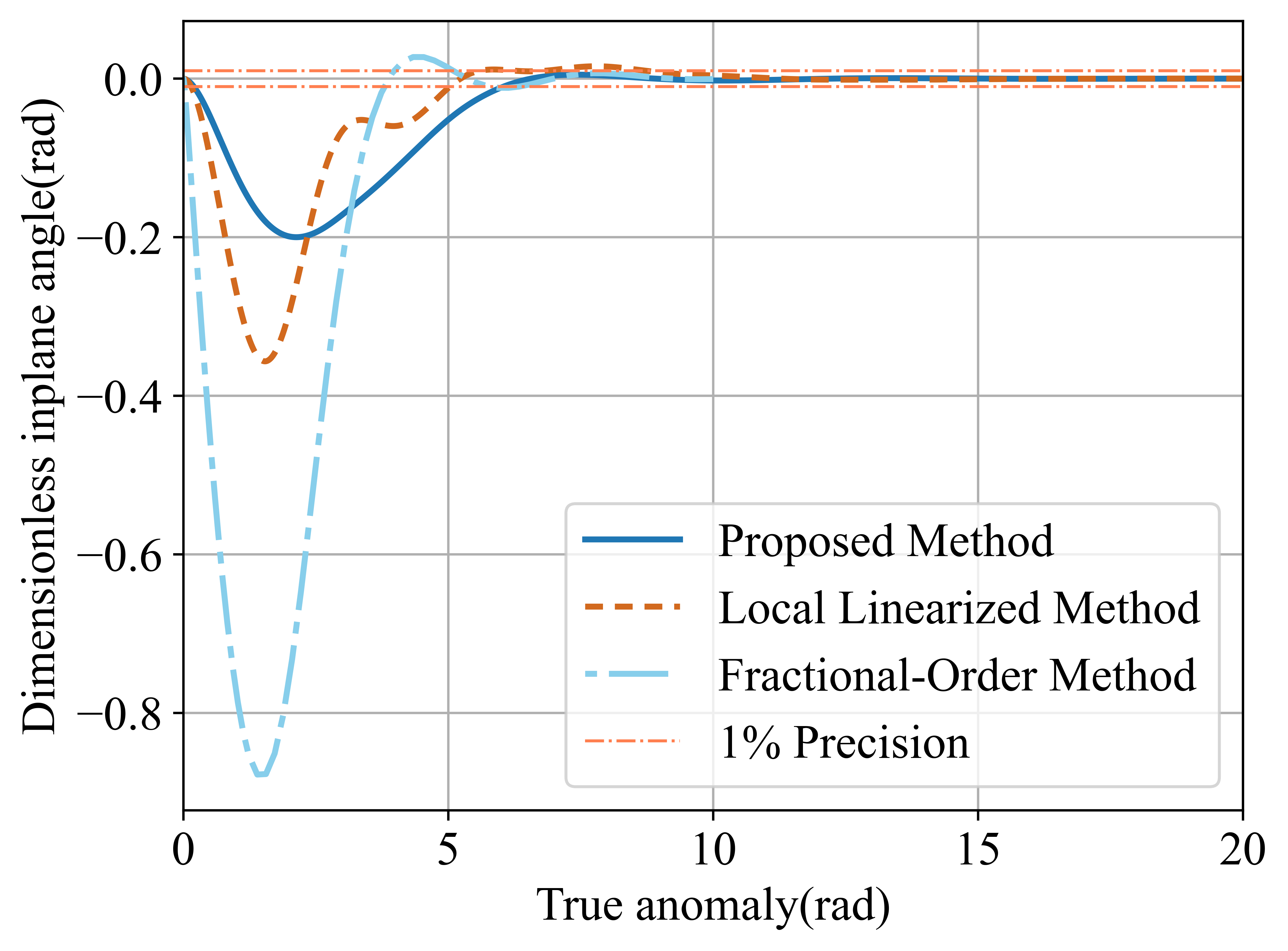} 
	\caption{Trajectory of in-plane angle} \label{fig6}
\end{figure}

The control law and parameters of the methods in \cite{Sun2014} and \cite{Pradeep1997} are chosen the same as that in their work. The LQR controller (\ref{lqr}) parameters are set to $K=36, Q'=I_{40},R'=15$ with two thousand simulation steps and sample period is specified as $h=0.01{\rm s}$. Simulation results are illustrated with Fig.\ref{fig5} and Fig.\ref{fig6}. It's generally thought that TSR deployment completes when the in-plane angle $x_1$ and tether length $x_3$ reached and remained the $1\%$ precision zone.

From the above plot and table, we verify the effectiveness of our proposed method, which uses a deep network to approximate a linear representation of a nonlinear system using the Koopman operator. All three methods complete the deployment around 10 $\rm{rad}$. And the proposed method has hardly any overshoot, which is more applicable to real deployment mission,  and completes deployment around 6 $\rm{rad}$ reaching $1\%$ precision zone more faster than the other two methods. The local linearized method \cite{Pradeep1997} deploys faster than the proposed method and fractional-order method at the beginning, but it has evident overshoot and takes a long time to eliminate it. The fractional-order method has little overshoot, but it takes longer time to achieve and stay $1\%$ precision zone. The fractional-order method can deploy more quickly than the proposed method at the beginning, which is an advantage in fast deployment missions. 

From Fig.\ref{fig6}, the proposed method's in-plane angle $\theta$ varies in $[-0.2,0]$, which is much smaller than the linearized method's $[-0.4,0]$ and the fractional-order method's $[-0.8,0]$. The faster deployment at the beginning causes the in-plane angle to change drastically in the linearized method and fractional-order method. The proposed method deploys a little slower than the other two methods, but the in-plane angle changes more gently and it completes deployment quicker with less overshot. It is easy to see that the proposed method owns better control performances than the other two methods. 

\begin{figure}[!t]
	\centering
	\includegraphics[width=18.5pc]{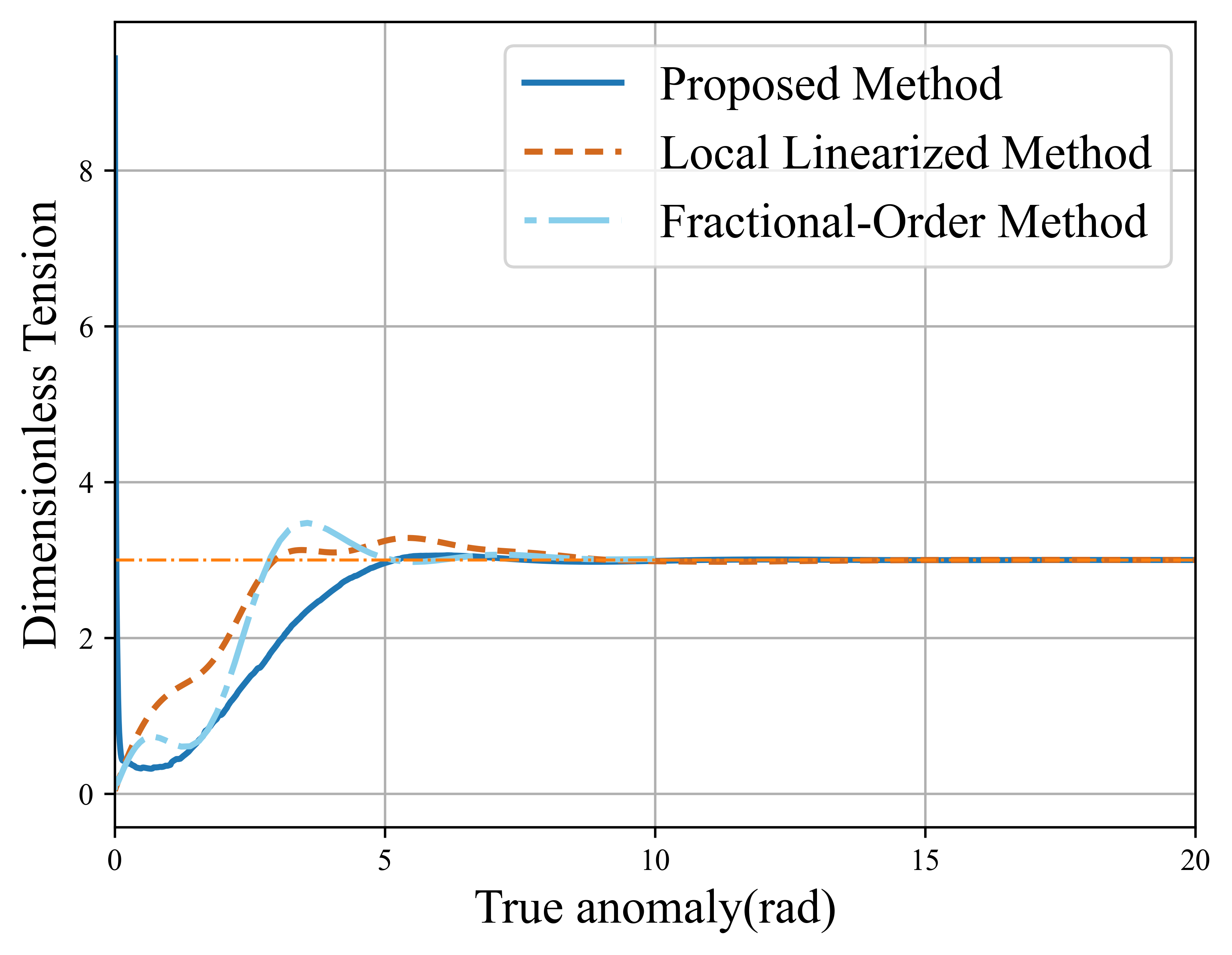} 
	\caption{Trajectory of dimensionless control input} \label{tension}
\end{figure}

Fig.\ref{tension} shows the simulation trajectory of $u$. It shows that the control input of proposed method varies more smoothly but it has an initial value and then decrease at the beginning. Recall the (\ref{opti_real_control}), the optimal control applied to TSR equals $\psi_{ux}(x_k)/\psi_{uu}(\hat{u}_k^*)$, and the denominator in our experiment is very close to zero at the beginning so that there is a non-zero control input. After that, the control input decreases to a small and positive value.

It is well known that there is a limited computation resource in the real space mission. Excessively complex algorithms cannot run on the onboard computer of satellite due to its limited computation ability. So we test the running time of three methods. The proposed method, the linearized method, and the fractional-order method are running 100 times with 2000 steps on the same computer, consuming 0.0122s, 0.00862s, 0.0821s respectively. Given the aforementioned, the proposed method possesses the better performance among the control ability and computation efficiency than the other two methods mentioned above in the real deployment missions.

\section{CONCLUSION}
This paper presents a data-driven optimal control framework for TSR deployment. The proposed method employs the deep learning based Koopman operator to convert the nonlinear TSR's dynamics into a global and equivalent linear system which is ready to apply linear control methods. As there is not a universal method to find the embedding functions $\Phi$ for a specific nonlinear system generally, a deep learning framework without decode net is employed to find them in this work. And the state and control matrix of lifted linear system in the embedding space are also estimated during training deep neural network in order to get a more robust approximation. The stability of lifted system and prediction ability of network are guaranteed with three designed loss functions. With the equivalent lifted linear system, linear optimal control methods can be applied to nonlinear dynamics like (\ref{state_dynamics}) directly without solving nonlinear optimal problem. Compared to other nonlinear control methods, the proposed method deploys TSR more quickly, smoothly and has hardly any overshoot with the linear optimal control policy. The proposed method deploys TSR a little bit slowly at the beginning and does not take the noisy data from sensor into consideration. In the future, our efforts will focus on solving these issues. In fact, the proposed data-driven optimal control framework is model-free, as it does not need any prior information about the system but merely the input and output data or trajectory of the system. The proposed method could extend to many other nonlinear systems that are difficult to model and design a stable controller.

\bibliographystyle{cas-model2-names}
\bibliography{main.bib, extra.bib}

\end{document}